\documentclass[]{llncs}
\usepackage{amsmath,mathptm}
\usepackage{amsfonts}
\usepackage{amssymb}
\usepackage{scharts}
\usepackage{color}
\usepackage{mathrsfs}
\usepackage{graphicx}
\title{Statechart Verification with iState}
\author{Dai Tri Man L\^e}
\institute{Department of Computer Science,\\
University of Toronto,\\
Toronto, ON, M5S 3G4 Canada\\
{ledt@cs.toronto.edu}}

\input{iCode.def}  
\begin{document}
\maketitle
\pagestyle{headings}

\begin{abstract} This paper is the long version of the extended abstract with the same name \cite{Le}. We describe in detail the algorithm to generate verification conditions from statechart structures implemented in the iState tool. This approach also suggests us a novel method to define a version of predicate semantics for statecharts analogous to how we  assign predicate semantics to programming languages.
\end{abstract}

\section{Introduction}
The statechart formalism, proposed by Harel \cite{Ha87} as an extension of conventional finite state machines, is a visual language for specifying reactive systems. It addresses the state explosion problem of state transition diagrams when modeling systems with parallel threads of control by introducing the concepts of \emph{hierarchy}, \emph{concurrency}, and \emph{communication}.

The \ISS tool translates statecharts into various programming languages, currently the Abstract Machine Notation (AMN) of the B method \cite{Abrial96:B}, Pascal, and Java. The translation is based on a definition of statecharts in terms of an extension of Dijkstra's guarded commands \cite{Sek1,Sek2}. This work demonstrates a novel statechart verification approach using \textit{state invariants} that has been added to \IS.

\section{Invariants}
Statecharts allow executable specifications to be derived from user requirements. We propose to supplement a statechart specification by \emph{invariants}. These are attached to states and specify what has to hold in a state configuration. Invariants are also derived from the requirements. They are not meant for execution, but they allow the statechart specification to be cross-checked. By themselves, statecharts do not lead to opportunities for consistency checks beyond well-formedness; invariants address this limitation and give a way of documenting the ``purpose'' of states.

Formally, invariants are predicates over global variables, like $x$ in the example below, and states (state tests):

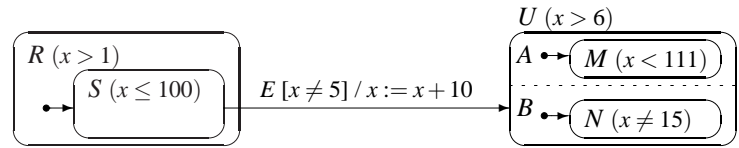
\begin{center}
\begin{statechart}(920,180)
\begin{andstate}(0,0){300}{150}
                \statename(30,110){$R$ \footnotesize $(x>1)$}

                \begin{basicstate}(80,10){200}{90}
                \statename(110,65){$S$ \footnotesize $(x \leq 100 )$}
                \end{basicstate}

                \initialstate(40,46)
                \begin{transition}(50,50)
                \fromstate(44,50)
                \tostate(80,50)
                \end{transition}

                \begin{transition}(280,50)
                \fromstate(280,50)
                \tostate(660,50)
                \transitioninfo[c](470,60){\footnotesize $E$ [$x \not = 5$] / 
                      $x:=x+10$}
                \end{transition}
\end{andstate}
                
\begin{andstate}(660,0){300}{150}
	 \statename(20,160){$U$ \footnotesize $(x>6)$}
	 
        \begin{xorstate}[n](0,80){300}{80}
        \statename(20,30){$A$}
            \begin{basicstate}(80,10){200}{50}
            \statename(110,25){$M$ \footnotesize $(x < 111)$}
            \end{basicstate}
            \initialstate(42,36)
            \begin{transition}(56,40)
                \fromstate(45,40)
                \tostate(78,40)
            \end{transition}
        \end{xorstate}
        \drawseparator(0,80)(300,80)
        \begin{xorstate}[n](0,0){300}{80}
        \statename(20,40){$B$}
            \begin{basicstate}(80,10){200}{50}
            \statename(110,25){$N$ \footnotesize $(x \not= 15)$}
            \end{basicstate}
            \initialstate(42,36)
            \begin{transition}(56,40)
                \fromstate(45,40)
                \tostate(78,40)
            \end{transition}
        \end{xorstate}
\end{andstate}

\end{statechart}
\end{center}

\noindent The definition of statecharts in \cite{Sek1,Sek2} translates states into variables and events into (nondeterministic) operations, in which use of the \emph{independent (parallel) composition} of statements is made; the parallel composition operator is essential for translating events with transitions in concurrent states. Using AMN, the states of the previous statechart are translated to variables $root \in \{R,U\}$, $r\in\{S\}$, $a\in\{M\}$ and $b\in\{N\}$ and the event $E$ is translated to:
\begin{eqnarray*}
E &\triangleq& \IF root = R \THEN \\
  &          & \quad\IF r = S \THEN \\
  &          & \quad\quad\IF x \neq 5 \THEN \\
  &          & \quad\quad\quad x:=x+10 ~\| ~root:= U~ \| ~a := M~ \| ~b := N \\
  &          & \quad\quad\END\\
  &          & \quad\END\\
  &          & \END
\end{eqnarray*}

\noindent Let $si$: \emph{State} $\to$ \emph{Condition} be a function that assigns to each state the invariant specified by the designer, or \emph{true} if none is specified, together with a test for being in that state.
For example:
\begin{eqnarray*}
si(S) & = & (r = S \wedge x \leq 100) \\
si(U) & = & (root = U \wedge x > 6)
\end{eqnarray*}
By the hierarchical structure of statechart, being in a state also means being in all of its ancestor states, in exactly one of its child states if the state is an XOR state, and in all of its child states if the state is an AND state. Hence, we have to compose state invariants together to create the \emph{accumulated invariant} $ai(s)$ of state $s$. For example:
\begin{eqnarray*}
  ai(S) & = & (root = R \wedge x > 1) \wedge (r = S \wedge x \leq 100) \\
  ai(U) & = & (root = U \wedge x > 6) \wedge
              ((a = M \wedge x < 111) \wedge (b = N \wedge x \neq 15))
\end{eqnarray*}
Formally, let \emph{Basic}, \emph{XOR}, \emph{AND} be disjoint subsets of the set \emph{State}. The accumulated invariant $ai$: \emph{State} $\to$ \emph{Condition} is defined with the help of the \emph{child invariant} $ci$: \emph{State} $\to$ \emph{Condition} as follows:
\begin{eqnarray*}
\CI(s) & \triangleq & {\begin{cases}
\SI(s) \wedge \bigotimes \CI[children[\{s\}]] & \text{ if } s\in \emph{XOR} \\
\SI(s) \wedge \bigwedge \CI[children[\{s\}]]     & \text{ if } s\in AND \\
\SI(s)                                           & \text{ if } s\in Basic
\end{cases}}
\\
\AI(s) & \triangleq & \bigwedge \SI[parent^+[\{s\}]] \wedge \CI(s)
\end{eqnarray*}
Here, $children[\{s\}]$ denotes the set of all child states of a state $s$ and $parent^+[\{s\}]$ denotes the set of all ancestor states of $s$, where $parent$ is the inverse of the $child$ relation, $parent = child^{-1}$ \cite{Sek1,Sek2}. The operator $\bigotimes$ stands for \emph{xor}. The definition reflects the meaning of XOR and AND states.

\section{Event Codes and Verification Tuples}

For each transition $E[guard]/action$ from state $S$ to $T$, where $action$ is a statement that may read and write to global variables, may include state tests, and may broadcast other events, a verification condition is generated: 
\[\{\AI(S) \wedge guard\} \emph{ action } \{\AI(T)\}\]
In the case of broadcasting in \emph{action}, the broadcast is replaced by a call to the corresponding operation. In the case of transitions in concurrent states on the same event $E$, a combined transition is considered. In the example, the verification condition for event $E$ is:
\begin{eqnarray*}
&& \{(\emph{root} = R \wedge x > 1) \wedge (r = S \wedge x \leq 100) \wedge x \neq 5 \} \\
&& ~~x:=x+10~\|~root:= U~\|~a := M~\|~b:= N \\
&& \{(root = U \wedge x > 6) \wedge
     ((a = M \wedge x < 111) \wedge (b = N \wedge x \neq 15))\}
\end{eqnarray*}
Hence, our goal is to automate the generating verification condition process. However, before doing so we need to have data structures to store and manipulate the verification conditions efficiently.

Using the algorithm discussed in \cite{Sek1,Sek2}, we map each statechart data structure into nondeterministic operations which is represented using an abstract syntax tree (AST). The AST of intermediate language is stored using the data type \emph{EventCode}
\[\emph{EventCode} \triangleq  \emph{Identifier}\pfun \emph{Statement}\]
where $S\pfun T$ denotes the set of partial function from $S$ to $T$.

Let $\PS{S}$ and $\seq{S}$ denote the types power set of $S$ and finite sequences of $S$ respectively. Also let $S \rel T$ denote the set of relation from $S$ to $T$. Each \emph{Statement} is then defined as a recursive data type
\begin{eqnarray*}
\emph{Statement} & \triangleq &  \emph{StateAssign}\quad\emph{State}\\
            &|&\emph{Assignment}\quad\emph{Identifier}\ \emph{Expression} \\
            &|&\emph{Bcast}\quad \emph{Identifier}\\ 
            &|&\emph{Guard}\quad \emph{Condition} \rel \emph{Statement}\\
            &|&\emph{Par}\quad \seq{\emph{Statement}}\\ 
            &|&\emph{Seq}\quad \seq{\emph{Statement}}\\
            &|&\emph{Skip}
\end{eqnarray*}
where:
\begin{itemize}
 \item $\emph{StateAssign}\quad\emph{State}$: denotes the state assignment node.
 \item $\emph{Assignment}\quad\emph{Identifier}\ \emph{Expression}$: denotes assignment node and the left hand side of the assignment is an identifier and the right hand side is an expression.
 \item $\emph{Bcast}\quad\emph{Identifier}$: denotes a broadcasting of an event whose name is represented using an identifier.
 \item $\emph{Guard}\quad\emph{Condition} \rel \emph{Statement}$:  denotes an \emph{alternative choice} where each choice is guarded using a condition. Notice that we use $\emph{Condition} \rel \emph{Statement}$ to emphasize the possible non-determinism. When several conditions are true at the same time, a choice is made non-deterministically.
 \item $\emph{Par}\quad\seq{\emph{Statement}}$: denotes the parallel composition of a sequence of statements. Due to the commutativity of parallel composition, we might use a set of statements instead of sequence. However, we decide to use sequence here only for the sake of determinism.
 \item $\emph{Seq}\quad\seq{\emph{Statement}}$: denotes the sequential composition of a \emph{sequence} of statements.
 \item $\emph{Skip}$ denotes the skip statement.
\end{itemize}

The type \emph{Condition} can either be a state test or a predicate, which is defined as following
\begin{eqnarray*}
\emph{Condition} &\triangleq & \emph{StateTest}\quad\emph{State}\\
            &|&\emph{Predicate}\indent\emph{Expression}
\end{eqnarray*}
Notice that using a functional language like syntax to define \emph{Statement} and \emph{Condition} allows us to employ pattern-matching on these data types when presenting our algorithms.

We then generate verification conditions from \emph{EventCode} data structure. This will be done by analyzing the structure of \emph{EventCode} to generate \emph{local verification conditions} and composing them together suitably to produce the final verification conditions. We treat all of these verification conditions uniformly using the notion of \emph{verification tuple}, which we think to be a more suitable representation of Hoare's triple for our verification purpose. The verification tuple type is define as following:
\[\emph{VTuple}  \triangleq   \PS{\emph{State}}\times\emph{Expression}\times\emph{Statement}\times\PS{\emph{State}}\]
where for each $(s,g,a,t) \in \emph{VTuple}$ we have:
\begin{itemize}
 \item $s$ denotes the set of source states of the transition,
 \item $g$ denotes the guard condition of the transition,
 \item $a$ denotes the statement which changes the states of global variables (including state variables),
 \item $t$ denotes the set of target states of the transition.
\end{itemize}
Notice that each $(s,g,a,t) \in \emph{VTuple}$ is converted into the following verification condition
\[\set{\bigwedge\AI[s] \wedge g}\ a\ \set{\bigwedge\AI[t]}\]
which is verified as a normal Hoare's triple. However, using sets of source of target states gives a more optimal way of composing verification conditions together. This also helps us avoiding redundancy when generating accumulating invariants due to states having common ancestors. In other words, we use the following more efficient way to calculate the acumulated invariant of a state set. Let the state set closure function $cl:\PS{\emph{State}}\fun\PS{\emph{State}}$ be defined as following
\[cl(ss) \triangleq \bigcup\set{parent^*[\set{s}]~|~s\in ss}\]
where $R^*$ denotes the reflexive transitive closure of the relation $R$. We define the set accumulated invariant function $\SAI:\PS{\emph{State}}\fun \emph{Condition}$ as follows:
\begin{eqnarray*}
\overline{\SI}(s,ss) & \triangleq & {\begin{cases}
\SI(s)  & \text{ if } s\in Basic\\
\SI(s)  & \text{ if } s \in AND \cup XOR ~\wedge~ \emph{children}(s) \in cl(ss) \\
\CI(s)	& \text{ otherwise}
\end{cases}}
\\
\SAI(ss) & \triangleq & \bigwedge \set{\overline{\SI}(s,ss)~|~s \in cl(ss)}
\end{eqnarray*}

For convenience, we let $\AST: \emph{String}\rightarrow \emph{Expression}\cup\emph{Statement}$ denote the mapping from a String to its AST. Hence, the verification condition discussed in our example previously can be expresses using the following \emph{VTuple}:
\[(\set{R,S}, ast(\text{``$x \neq 5$''}),ast(\text{``$x:=x+10~\|~root:= U~\|~a := M~\|~b:= N$''}),\set{U,M,N})\]

\section{Invariant Verification Algorithm}
Before presenting the algorithm for generating verification tuples, we need several auxiliary functions. We divide \emph{Condition} into \emph{StateTest} and \emph{Predicate}. The state test condition \emph{StateTest} are generated by EventCode generators using the algorithm in \cite{Sek1,Sek2}. Hence, we let \emph{Predicate} denote the transition guards supplied by users. To distinct these two cases, we use the following function\\
$
\\
\indent \emph{c2tuple}~:~\emph{Condition}\fun \emph{VTuple}\\
\indent \emph{c2tuple}\quad\emph{StateTest}\ s=(\set{s},\emph{true},Skip,\emptyset)\\
\indent \emph{c2tuple}\quad\emph{Predicate}\ e=(\emptyset,e,Skip,\emptyset)\\
\\
$
Considering the following example where you have:
\[\IF c_1 \THEN s_1 \ELSE s_2 \END \|\IF c_2 \THEN s_3 \ELSE s_4 \END\]
For simplicity, we suppose these two statements are simply user statements without state test or state assignment. Then the statement $\IF c_1 \THEN s_1 \ELSE s_2 \END$ corresponds to the verification tuple set
\[\set{(\emptyset,c_1,s_1,\emptyset),(\emptyset,\neg c_1,s_2,\emptyset)}\]
and the statement $\IF c_2 \THEN s_3 \ELSE s_4 \END$ corresponds to the set
\[\set{(\emptyset,c_2,s_3,\emptyset),(\emptyset,\neg c_2,s_4,\emptyset)}\]
Since these two if statements are composed in parallel, we the resulted parallel product verification tuple set for the whole statement is:
\begin{center}
$\set{(\emptyset,c_1\wedge c_2,s_1\|s_3,\emptyset),(\emptyset,\neg c_1 \wedge c_2 ,s_2\|s_3,\emptyset),
(\emptyset,\neg c_1\wedge c_2,s_2\|s_3,\emptyset),$\\
$(\emptyset,\neg c_1\wedge \neg c_2,s_2\|s_4,\emptyset)}$
\end{center}
The more general case including state tests are tackled using the following functions:\\
$
\\
\indent \emph{parProd}~:~\seq{\PS{\emph{VTuple}}}\fun\PS{\emph{VTuple}}\\
\indent \emph{parProd}\quad[] = \emptyset\\
\indent \emph{parProd}\quad[s] = s\\
\indent \emph{parProd}\quad[s_1,\ldots,s_n] = \set{concat1([t_1,\ldots,t_n])~|~(t_1,\ldots,t_n) \in s_1\times\ldots\times s_n}\\
\\
\indent \emph{concat1}~:~\seq{\emph{VTuple}}\fun \emph{VTuple}\\
\indent \emph{concat1}\quad[] = (\emptyset,\empty{true},\emph{Skip},\emptyset)\\
\indent \emph{concat1}\quad[(s_i,g_i,a_i,t_i)~|~i=1..n] = (\bigcup_{i=1}^{n} s_i,\bigwedge_{i=1}^{n} g_i,\emph{Par}\ [a_1,\ldots,a_n],\bigcup_{i=1}^{n} t_i)\\
\\
$
This function \emph{parProd} is used very often in our implementation and it helps to simplify the implementation substantially.

Similar to the case of parallel composition is the case sequential composition. For example, an event code
\[\IF c_1 \THEN s_1 \ELSE s_2 \END~;~\IF c_2 \THEN s_3 \ELSE s_4 \END\]
will be translated into the following verification tuple set
\begin{center}
$\set{(\emptyset,c_1\wedge c_2,s_1;s_3,\emptyset),(\emptyset,\neg c_1 \wedge c_2 ,s_2;s_3,\emptyset),
(\emptyset,\neg c_1\wedge c_2,s_2;s_3,\emptyset),$\\
$(\emptyset,\neg c_1\wedge \neg c_2,s_2;s_4,\emptyset)}$
\end{center}
Since the statements must be composed sequentially, the functions are implemented as following.
$
\\
\indent \emph{seqProd}~:~\seq{\PS{\emph{VTuple}}}\fun\PS{\emph{VTuple}}\\
\indent \emph{seqProd}\quad[] = \emptyset\\
\indent \emph{seqProd}\quad[s] = s\\
\indent \emph{seqProd}\quad[s_1,\ldots,s_n] = \set{concat2([t_1,\ldots,t_n])~|~(t_1,\ldots,t_n) \in s_1\times\ldots\times s_n}\\
\\
\indent \emph{concat2}~:~\seq{\emph{VTuple}}\fun \emph{VTuple}\\
\indent \emph{concat2}\quad[] = (\emptyset,\empty{true},\emph{Skip},\emptyset)\\
\indent \emph{concat2}\quad[(s_i,g_i,a_i,t_i)~|~i=1..n] = (\bigcup_{i=1}^{n} s_i,\bigwedge_{i=1}^{n} g_i,\emph{Seq}\ [a_1,\ldots,a_n],\bigcup_{i=1}^{n} t_i)\\
$

We then define a function $\emph{s2tuples}$ which maps each statement node to a set of verification tuples. Using pattern-matching, the function \emph{s2tuples} can be defined as following:\\
$
\\
\indent\emph{s2tuples}~:~\emph{Statement}\fun\PS{\emph{VTuple}}\\
\indent\emph{s2tuples}\quad\emph{StateAssign}\ s = \set{(\emptyset,\emph{rue},\emph{StateAssign}\ s,\set{s})} \\
\indent\emph{s2tuples}\quad\emph{Assignment}\ i\ e = \set{(\emptyset,\emph{true},\emph{Assignment}\ i\ e,\emptyset)} \\
\indent\emph{s2tuples}\quad\emph{Bcast}\ i = \set{(\emptyset,\emph{true},\emph{Bcast}\ i,\emptyset)} \\
\indent\emph{s2tuples}\quad\emph{Guard}\ r = \bigcup \set{parProd([\set{\emph{c2tuple}(c)},\emph{s2tuples}(s)])~|~(c,s)\in r} \\
\indent\emph{s2tuples}\quad\emph{Par}\ ss = \emph{parProd}([\emph{s2tuples}(s)~|~s\leftarrow ss])\\
\indent\emph{s2tuples}\quad\emph{Seq}\ ss = \emph{seqProd}([\emph{s2tuples}(s)~|~s\leftarrow ss])\\
\indent\emph{s2tuples}\quad\emph{Skip} = (\emptyset,\emph{true},Skip,\emptyset)\\
$

Hence, we can construct a global verification tuple map for all events using the following function:\\
$
\\
\indent\emph{vtupleMap}~:~\emph{EventCode}\fun(\emph{Identifier}\fun\PS{\emph{VTuple}})\\
\indent\emph{vtupleMap}\ ec= \emph{s2tuples}\circ ec \\
$\\
where the operation $\circ$ denotes the usual function composition, i.e, $f\circ g (x) \triangleq f(g(x))$.

Since each verification tuple $(s,g,s,t) \in \bigcup ran(vtupleMap)$, the action $s$ might still contain event broadcasting. We deal with these event broadcasts similarly to the case of parallel composition of event codes. We first apply topological sorting algorithm \cite{Algo} on $\emph{vtupleMap}(ec)$ to obtain a sequence 
\[tss = [(e_1,s_1),\ldots,(e_n,s_n)] \in \seq{\emph{Identifier}\times\PS{\emph{VTuple}}}\]
such that for each $(e_i,s_i)$, the verification tuple set $s_i$ contains the actions that boardcast only the events in the set $\set{e_1,\ldots,e_{i-1}}$. We can always obtain such a list with the assumption that we don't allow circular broadcasting \cite{Beck94}. 
We next define a function to collect and filter the boardcasts from the action of a verification tuple as following:\\
$
\\
\indent\emph{collectBcast}~:~\emph{Statement}\fun\PS{\emph{Identifier}}\\
\indent\emph{collectBcast}\quad\emph{Bcast}\ i = \set{i}\\
\indent\emph{collectBcast}\quad\emph{Guard}\ \set{(c_i,s_i)~|~i=1..n} = \bigcup_{i=1}^{n}\emph{collectBcast}(s_i)\\
\indent\emph{collectBcast}\quad\emph{Par}\ [s_1,\ldots,s_n] = \bigcup_{i=1}^{n}\emph{collectBcast}(s_i)\\
\indent\emph{collectBcast}\quad\emph{Seq}\ [s_1,\ldots,s_n] = \bigcup_{i=1}^{n}\emph{collectBcast}(s_i)\\
\indent\emph{collectBcast}\quad\emph{s} = \emptyset\\
\\
\indent\emph{filterBcast}~:~\emph{Statement}\fun\emph{Statement}\\
\indent\emph{filterBcast}\quad\emph{Bcast}\ i = \emph{Skip}\\
\indent\emph{filterBcast}\quad\emph{Guard}\ r = \emph{Guard}\ \set{(c,s)~|~(c,s)\in r \wedge \neg \emph{isBcast}(s)}\\
\indent\emph{filterBcast}\quad\emph{Par}\ ss = \emph{Par}\ [s~|~s\leftarrow ss \wedge \neg \emph{isBcast}(s)]\\
\indent\emph{filterBcast}\quad\emph{Seq}\ ss = \emph{Seq}\ [s~|~s\leftarrow ss \wedge \neg \emph{isBcast}(s)]\\
\indent\emph{filterBcast}\quad\emph{s} = s\\
\\
\indent\emph{isBcast}~:~\emph{Statement}\fun\emph{Boolean}\\
\indent\emph{isBcast}\quad\emph{Bcast}\ \_~=~true \\
\indent\emph{isBcast}\quad\_~=~false\\
$

In our real implementation, we filter and collect broadcasted events at the same time for the sake of efficiency. We let $\smallfrown$ denote the sequence concatenation operator. Then the process of translating verification tuples with broadcasting to ones without boardcasting is implemented as the following functions:\\
$
\\
\indent\emph{vtupleNoBcast}~:~(\emph{VTuple}\times \seq{\emph{Identifier}\times\PS{\emph{VTuple}}})\fun\PS{\emph{VTuple}}\\
\indent\emph{vtupleNoBcast}\quad((s,g,a,t)~,~\emph{tspre}) = \\ 
\indent\indent\indent\IF (\emph{collectBcast} = \emptyset) \THEN\\
\indent\indent\indent\indent\emph{parProd}([\set{(s,g,\emph{filterBcast}(a),t)}]^\smallfrown\\
\indent\indent\indent\indent\indent\indent\indent\indent[s_i~|~(e_i,s_i)\leftarrow \emph{tspre}\wedge e_i\in \emph{collectBcast}(a)])\\
\indent\indent\indent\ELSE\\
\indent\indent\indent\indent\set{(s,g,a,t)}\\
\indent\indent\indent\END\\
$\\
The function \emph{vtupleNoBcast} takes a pair of verification tuples and \emph{tspre} as inputs. The argument \emph{tspre} represents the set of prefix of $tss$ where all boardcasting are already expanded. Since we $tss$ is already topologically sorted according to the dependencies of boardcasting, the action $a$ only broadcasts the events defined in \emph{tspre}. Hence, we apply the function $\emph{parProd}$ on the list consisting of the input verification tuples with all the broadcasts filtered and the part of \emph{tspre} chosen according the set of events that the action $a$ broadcasts. 

The rest of the elimination of broadcasting is defined in the next two functions. We define a function \emph{vtupleSetNoBcast} which is similar to \emph{vtupleNoBcast} but apply on verification tuple sets instead.\\
$
\\
\indent\emph{vtupleSetNoBcast}~:~(\PS{\emph{VTuple}}\times \seq{\emph{Identifier}\times\PS{\emph{VTuple}}})\fun\PS{\emph{VTuple}}\\
\indent\emph{vtupleSetNoBcast}\quad (\set{v_i~|~i=1..n}~,~\emph{tspre})~=~\bigcup_{i=1}^{n}\emph{vtupleNoBcast}(v_i,\emph{tspre})\\
\\
$
We then define the desired function \emph{vseqNoBcast} which can now be applied to the topological sort list $tss$ to expand the event broadcasting to suitable verification tuple sets. This function is defined as following:\\
$
\\
\indent\emph{vseqNoBcast}~:~ \seq{\emph{Identifier}\times\PS{\emph{VTuple}}}\fun\seq{\emph{Identifier}\times\PS{\emph{VTuple}}}\\
\indent\emph{vseqNoBcast}\quad[]~=~[]\\
\indent\emph{vseqNoBcast}\quad[(e_i,s_i)~|~i=1..n]~=~\emph{tspre}^\smallfrown[(e_n,\emph{vtupleSetNoBcast}(s_n,\emph{tspre}))]\\
\indent\indent\indent\WHERE\quad\emph{tspre}~=~\emph{vseqNoBcast}([(e_i,s_i)~|~i=1..n-1])\\
\\
$
Hence, we use the result $\emph{vseqNoBcast}(tss)$ to generate the verification conditions.

\section{Predicate Semantics of Statecharts}
Our verification approach reveals a strong connection between a statechart transition and a verification tuple. This motivates us to provide a predicate semantic of statecharts instead of the traditional operational way in \cite{HarelNaamad96,QinChin,Andrea}. We do so by introducing the functions for translating statecharts to verification tuples. These functions are very similar to the functions used in the previous section. In fact, we will use some auxiliary functions defined previously. 

We first need to provide the data structure used to represent statecharts.
\begin{eqnarray*}
\emph{State} & \triangleq &  \emph{Basic}\quad \emph{Identifier}\\
            &|&\emph{And}\quad\emph{Identifier}\times \seq{State} \\
            &|&\emph{Xor}\quad\emph{Identifier}\times \seq{State}\times\emph{State}\times\emph{Transition}
\end{eqnarray*}
where \emph{Transition} is a five-ary relation defined as following:
\[\emph{Transition} \triangleq \emph{State}\times \emph{Identifier} \times\emph{Condition}\times \emph{Statement}\times\emph{State}\]
This definition says a statechart state can be either
\begin{itemize}
 \item a Basic state with a state name,
 \item a composite And state encoded by a state name and a sequence of parallel sub states, or
 \item a composite Xor state encoded by a state name, a sequence of sub states, an initial state and a transition relation. 
\end{itemize}
Each transition is defined by its source state, triggering event, transition guard, action and target state respectively.

We also use an event-centric approach by dealing with each event separately. Hence, we first define a function to return the restriction of a statechart with respect to a specific event. We use $\emps$ to denote the empty identifier which indicate the ``event name'' of a spontaneous transitions.\\
$
\\
\indent\emph{resStateEvent}~:~ (\emph{Identifier}\times\emph{Sate})\fun\emph{State}\\
\indent\emph{resStateEvent}\quad(\_~,~\emph{Basic}\ id\ s)~=~\emph{Basic}\ id\ s\\
\indent\emph{resStateEvent}\quad(e~,~\emph{And}\ id\ [s_i~|~i=1..n])~=~\emph{And}\ id\ [\emph{resStateEvent}(i,s_i)~|~i=1..n]\\
\indent\emph{resStateEvent}\quad(e~,~\emph{Xor}~~id~~seqs~~init~~t)\\
\indent\indent\indent~=~\emph{Xor}~~id~~seqs~~init~~[(ss,e',c,a,ts)~|~(ss,e',c,a,ts)\leftarrow t ~\wedge~ (e'=e \vee e'=~\emps)]\\
\\
$
We then need a function to return the verification tuple correspondent to the initialization of a state due to the fact that each composite XOR state must have an initial state.\\
$
\\ 
\indent\emph{initialize}~:~\emph{State}\fun\PS{\emph{Vtuple}}\\
\indent\emph{initialize}\quad s ~=~ \CASE~ s ~\OF\\
\indent\indent\emph{Basic}~~id~~s~\rightarrow~\set{(\emptyset,\emph{true},\emph{StateAssign}\ s,\set{s})}\\
\indent\indent\emph{And}~~id~~[s_i~|~i=1..n]~\rightarrow\\
\indent\indent\indent\emph{parProd}([\emph{initialize}(s_i)~|~i=1..n]^\smallfrown[\set{(\emptyset,\emph{true},\emph{StateAssign}\ s,\set{s})}])\\
\indent\indent\emph{Xor}~~id~~[s_i~|~i=1..n]~~init~~t~\rightarrow~\\
\indent\indent\indent\emph{parProd}([\emph{initialize}(init)]^\smallfrown[\set{(\emptyset,\emph{true},\emph{StateAssign}\ s,\set{s})}])\\
\\
$
This function \emph{initialize} will always return a \emph{singleton set}, since we enforce only one possibility to initialize a composite or basic state by disallowing some statecharts variants \cite{Sek1,Sek2}. However, for convenience the returned type of this function is $\PS{\emph{Vtuple}}$ to make it easier for composing verification tuples together in the intermediate composition steps. In other words, it allows us to treat this case as a special case of parallel composition using the \emph{parProd} function.

The next step is to define a recursive function to generate verification tuples with respect to statechart data structure. The base case is the the case of basic states and the recursive cases deal with composite states. The most difficult problem is caused by the existence of spontaneous transitions in composite XOR states. We define the following two functions that take a child state of an XOR state and generate verification conditions. When there are spontaneous transitions, the function \emph{getNext} invokes \emph{getSpon} to search for the spontaneous transition going from the target states and generate verification condition for them and then compose the verification conditions together properly.\\ 
$
\\
\indent\emph{getNext}~:~(\emph{State}\times\emph{Transition})\fun\PS{\emph{Vtuple}}\\
\indent\emph{getNext}\quad(s,tr)\\
\indent\indent=~ \set{\emph{parProd}([\set{(\set{s},g,a,\emptyset)},\emph{getSpon}(t,tr)])~|(s,id,g,a,t)\in tr ~\wedge~ id \not=~ \emps}\\
\\
\indent\emph{getSpon}~:~(\emph{State}\times\emph{Transition})\fun\PS{\emph{Vtuple}}\\
\indent\emph{getSpon}\quad(s,tr)~=~S \cup \emph{parProd}([\set{(\emptyset,G,Skip,\emptyset)},\emph{initialize}(s)])\\
\indent\indent\WHERE\\
\indent\indent\indent S = \set{\emph{parProd}([\set{(\emptyset,g,a,\emptyset)},\emph{getSpon}(t,tr)])~|(s,id,g,a,t)\in tr ~\wedge~ id =~ \emps}\\
\indent\indent\indent G = \bigwedge\set{\neg g~|~(s,id,g,\_,\_)\in tr ~\wedge~ id = \emps}\\
\\
$
The set $S$ in function \emph{getSpon} corresponds to the case the spontaneous transitions are taken and the condition $G$ is the condition for non of the spontaneous transition from state $s$ is taken.

After having these two functions, the rest of the task of generating verification tuples from a statechart state \emph{restricted to one event} is defined in the following function.\\
$
\\
\indent\emph{s2tuples}~:~\emph{State}\fun\PS{\emph{Vtuple}}\\
\indent\emph{s2tuples}\quad s ~=~ \CASE~ s ~\OF\\
\indent\indent\emph{Basic}~~id~~s~\rightarrow~\set{(\set{s},\emph{true},\emph{Skip},\emptyset)}\\
\indent\indent\emph{And}~~id~~[s_i~|~i=1..n]~\rightarrow~\emph{parProd}([\emph{s2tuples}(s_i)~|~i=1..n]])\\
\indent\indent\emph{Xor}~~id~~[s_i~|~i=1..n]~~init~~tr~\rightarrow~S_1 \cup S_2\\
\indent\indent\indent\WHERE\\
\indent\indent\indent\indent S_1=\bigcup_{i=1}^n\emph{getNext}(s_i,tr)\\
\indent\indent\indent\indent G=\bigwedge\set{\neg g~|~(\_,g,\_,\_)\in S}\\
\indent\indent\indent\indent S_2=\emph{parProd}([\set{(\emptyset,G,Skip,\emptyset)}]^\smallfrown[\emph{s2tuples}(s_i)|i=1..n])\\
\\
$
To generate all the verification conditions of a statechart, we first collect the set of all events in a given statechart. For each event we use the function \emph{resStateEvent} to get the restricted statechart to each event in the set and then apply the function \emph{s2tuples} to the root of the statecharts. As a result of this process, we will get a map from event names to verification tuples exactly like the map \emph{vtupleMap} previously. The branch statements and event broadcasting in the actions of verification tuples can be easily expanded out using the function \emph{s2tuples} and \emph{vseqNoBcast} defined in the last section.

\section{Implementation}

The \ISS tool currently uses the Simplify theorem prover \cite{Rus} to discharge the generated verification conditions because of its support of first order logic and linear arithmetic. Simplify also has arrays built in, though currently \ISS does not use them. We are working on extending \ISS with data types like arrays, rational numbers, and real numbers. In future, we also plan to extend the verification theory to timed transitions \cite{Saeed}. 

\section{Discussion}
\enlargethispage*{0.1cm}
Compared to the statechart verification approaches in \cite{Sur,Cla,Mikk}, we use an \textit{event-centric} semantics of statecharts by looking at events as \textit{operations} rather than \textit{data} as in the original \textit{state-centric} semantics \cite{HarelNaamad96}. Instead of writing global temporal specification (say in CTL or LTL) separately, inspired by \textit{nested invariant diagram} \cite{Back}, invariants (\emph{safety properties}) are attached to states.

By attaching invariants to states and utilizing the guarded command representation of statecharts \cite{Sek1,Sek2}, we arrive at a rather straightforward verification method. The approach generating verification conditions leads to many small ``local'' verification conditions and avoids some impossible configurations, compared to when specifying invariants on the global level. As many small verification conditions are easier to handle automatically than a few large ones, we believe that the approach can more easily scale up for the verification of large systems.

\bibliographystyle{plain}
\bibliography{istate}

\end{document}